\documentclass[preprint,12pt]{elsarticle}

\usepackage{amssymb}

\begin{document}
\begin{frontmatter}



\title{Modelling the impact of financialization on agricultural commodity
markets}


\author[SISSA]{Maria d'Errico}

\author[SISSA]{A.~Laio}

\author[SISSA,CNR,IMT]{G.L.~Chiarotti\corref{cor2}}

\address[SISSA]{International School for Advanced Studies, Via Bonomea 265, IT-34151
Trieste}

\address[CNR]{Istituto dei Sistemi Complessi, ISC-CNR, Via dei Taurini 19, IT-00185
Roma}

\address[IMT]{Scuola IMT Alti Studi Lucca, Piazza S. Ponziano 6, IT-55100 Lucca}

\cortext[cor2]{Corresponding author: guido.chiarotti@gmail.com}

\begin{abstract}
We propose a stylized model of production and exchange in which long-term
investors set their production decision over a horizon $\tau$, the
``time to produce'', and are liquidity constrained, while financial
investors trade over a much shorter horizon $\delta$ ($\ll\tau$)
and are therefore more duly informed on the exogenous shocks affecting
the production output. The equilibrium solution proves that: 
\mbox{(i) long-term} producers modify their production decisions to anticipate the impact
of short-term investors allocations on prices; (ii) short-term investments
return a positive expected profit commensurate to the informational
advantage. While the presence of financial investors improves the
efficiency of risk allocation in the short-term and reduces price
volatility, the model shows that the aggregate effect of commodity
market financialization results in rising the volatility of both farms'
default risk and production output.
\end{abstract}



\end{frontmatter}



\section{Introduction}

\label{intro}

The integration of agricultural commodity and financial markets has
been largely criticized by both a growing body of economic literature
(see e.g. \cite{Orhangazi2008}) and political consideration \cite{Soros2008}.
According to those critics the overflow of capital invested in the
commodity markets feeds a destabilizing speculation, and a tightening
of market regulation is then claimed. While the empirical evidence
of international food price spikes and volatility driven by financial
markets is robust (see \cite{Tadesse2013} and references therein),
its mechanisms and potential drawbacks for production risk management
are unexplored.

This paper analyzes the effects of market integration on conventional
production planning and farm liquidity risk management policies within
a sty\-li\-zed equilibrium model of production, trade and consumption
of an agricultural commodity where financial and agricultural commodity
markets are partially integrated. We suggest that the \textquotedblleft time
to produce\textquotedblright\ is the fundamental parameter governing risk
allocation and production dynamics and should be considered in designing
more efficient production schemes and distribution policies aimed
at improving liquidity risk-sharing.

The stylized model assumes that the production of an agricultural
commodity is risky because farmers' resources allocation and production
decisions are irreversible and taken over a horizon $\tau$, the ``time
to produce''. On the other hand financial investors trade over a
much shorter horizon $\delta$ ($\ll\tau$) and possess an information
advantage contingent on the exogenous shocks affecting production
of the commodity. 
In our model, investments on contingent contracts play a role similar
to investments on physical storage and the zero net supply condition
is assumed to hold on average.

Actually, as discussed by \cite{Mello2013}, any hedging policy,
by requiring margin capital and liquidity, will in general have adverse
effects on liquidity needs. In particular the inefficiency of long-term
industry's risk management are documented and discussed in \cite{Parsons2011},
in \cite{Cooper1999}, \cite{Allayannis1999} and in \cite{Mello2000}.

Motivated by the above remarks, in the model at hand it is assumed
that long-term investors face a liquidity constraint: at the end of
the production cycle the farm goes bankrupt and its production is
lost if the profit resulting from selling the production at market
price is not sufficient to break even. Consequently, while price levels
continue to depend on the joint allocation decisions of financial
investors and producers, profits from financial hedging cannot be
used to soften the liquidity constraint faced by long-term investors.
In practice this constraint reproduces a segmentation of the capital
market which is empirically well documented: most of the profits in
commodity market trading remunerate professional investors while farm
hedging investments are low.

In the model the commodity equilibrium price is thus set by the clearing
of the market in an economy populated by an heterogeneous set of producers
exposed to idiosyncratic and systematic shock components, by consumers
described by an exogenously specified demand curve and by an extra
demand or supply contribution generated by financial agents.

A comparative statics exercise proves that this model can reproduce
many interesting stylized facts. First of all it is possible to prove
that the long-term investors modify the production choice due to the
effects of financial trading. In addition the financial investors
make profit by exploiting their informational advantage. Finally,
the model shows a progressive increase in the volatility of the produced
quantity and a growth in default risk with an increasing market integration.

Our model highlights that the major issue in farm risk management
is the necessity to alleviate the effects of credit constraints to
reduce price pressure on producers and consumers. Correspondingly,
public subsidization of agricultural investments can be interpreted
as a necessary (possibly inefficient) response to restore the balance
of capital flows from short-term to long-term investors. A rationalization
for the farm policy actions similar to our has been discussed also
in \cite{Leathers01111986}. Our approach, however, is more focused
on the description of farmers' production planning decisions and our
findings indicate that a rational production planning and liquidity
risk management must classify different productions and financing
opportunities considering the \textquotedblleft time to produce\textquotedblright\
constraint faced by different participants.

The paper is organized as follows: in Section \ref{mat} we define the model and discuss
 its innovations with respect to the past literature. In Section \ref{calc}
we present and discuss numerical results and their dependence on exogenous
parameters. Section \ref{res} is devoted to a discussion about
the assumptions made within the model, while Section \ref{conc} to
conclusions. A simplified (analytical solvable) version of the model
is provided in the Appendix, together with some techinical aspects
of the self consistent calculation procedure adopted.

\section{Materials and Methods}

\label{mat}

\subsection{Related literature}

\label{lit}

The model at hand is based on many simplifying assumptions in common
with other rational expectations competitive storage models, \cite{Brennan1958},
\cite{Williams1991}, \cite{Deaton-Laroque,Deaton1996} and \cite{Routledge2000}.

The intrinsic information asymmetry introduced in the model (discussed
in Section~\ref{model}) can be seen as a reduced form of the long-standing
hedging pressure theory of commodity prices that dates back to \cite{Keynes1930},
\cite{Hicks1939} and, more recently, to \cite{Hirshleifer1988}.

Market efficiency and its implications on commodity futures prices
dynamics dates back to the seminal contribution of \cite{Samuelson1971}.
In \mbox{presence} of perfect (infinitely liquid and frictionless) capital markets,
farmers could borrow sufficient capital to invest in the production
and simultaneously take a position in the contingent contract markets
to hedge the production risk.

The relation between liquidity constraints and productivity in farming
and agricultural industries is grounded on a vast literature, see
e.g. \cite{Griggeman-Towe-Morehart} and references therein. The
influence of biological production lags on agricultural commodity
price dynamics has been investigated in literature by means of the
well known cobweb model, for example in the case of U.S. beef markets
(see \cite{Chavas2000}). A thorough analysis of the relation between
credit and liquidity constraints, farm investment policies and optimal
subsidization policies could be found in \cite{Vercammen}.

In the equilibrium analysis described in \cite{Zhou1998} it is shown
that the presence of producers' liquidity constraints induces mean
reversion in futures prices, while government price subsidies, if
actively hedged by producers, lowers futures risk premia and reduces
price volatility. Differently from that approach, here we explicitly
model a multiplicity of producers which can default due to price dynamics.
In this way we can analyze the equilibrium relations between liquidity
restrictions, producers' default and commodity prices. In this respect,
our model bears some similarities with the model discussed in \cite{Acharya2009}
where limits to arbitrage generate limits to hedging by producers.

\subsection{The model}

\label{model}

Production, trade and consumption of a real good are explicitly modeled.
The price is set by the equilibrium between the supply and demand
when the market clears. In particular, the equilibrium price is determined
by the agents' interaction, their operational timeline and market
clearing conditions.

Hereafter agents are idealized to the extent that farmers can only
produce real goods and financial investors can only trade goods whose
payoff is contingent to the production outcome of the real one. The
two types of agents are considered separated from one another in order
to study the corresponding sector's returns. However, a real agent
could behave as a combination of the two idealized ones. As will be
commented below considering a real agent (for example a farmer that
can both produce and trade) does not alter the results of the model.

\subsubsection{Agents}

Supply and demand are regulated by the interaction of three types
of agents: farmers, consumers and financial investors.
\begin{itemize}
\item \textbf{Farmers} are the producers of the commodity goods in the economy.
We consider an ensemble of $N$ farmers, all producing the same food commodity,
which only are allowed to \textit{sell} the produced goods. A single
farmer produces a quantity $q$ of this commodity investing an amount
of capital $m$. We consider that each farmer incurs in a fixed cost
which do not depend on the quantity produced. For the sake of simplicity
we assume $c_{F}$ to be the same for all the farmers.

The operational decision of the quantity to be produced is taken by
the farmer at time $0$. The quantity $q$ will be available on the
market at a later time $\tau$, corresponding to the ``time to produce'',
and sold at a price $p_{\tau}$. Thus, the profit of a farmer will
be: 
\begin{equation}
\pi_{F}=p_{\tau}q_{\tau}-m-c_{F}\label{eq:p1}
\end{equation}

Producers are assumed to be risk neutral and, for the sake of simplicity,
a zero rate discounting is applied. Extensions of the model to both fixed cost farmers' dependency and to finite rate discounting
are straightforward and will be discussed elsewhere. Within
our framework the effect of individual risk aversion would be mitigated by the existence of a multitude of heterogeneous producers.

If the actual profit reaped by the farmer is negative the farmer defaults
and the amount produced is distributed among lenders and does not
contribute to the total quantity of goods brought to the market at
time $\tau$. Notice that 
in our model the liquidity constraint, that, as we will see is crucial
in determining the production output volatility, is introduced by
modeling the farmers\rq{} default. 
The quantity $q_{\tau}$ is determined by the investment level $m$, via
a production function $q_{\tau}\left(m\right)$, that is generically
assumed to be concave and with a positive derivative. We choose a
simple form of $q_{\tau}\left(m\right)$: 
\begin{equation}
q_{\tau}(m)=\theta_{\tau}\sqrt{m}\label{eq:prod}
\end{equation}
where $\theta_{\tau}$ is the 
\textquotedblleft fitness\textquotedblright \ which induces uncertainty in the final
amount specific to each farmer. $\theta$ models the exogenous stochastic
uncertainty shocks and its realization at time $\tau$ is not predictable
by farmers at the time of the production operational decision. However,
its probability distribution (described in Section~\ref{2.2}) is
assumed to be known.

The time interval $\tau$ represents the lag between the farmers'
operational investment and the market clearing epoch (in short, the
``time to produce'').

All the farmers enter the market at time $t=0$ in exactly the same
conditions \mbox{$\theta=\theta_{0}$}. Each farmer sets an investment level $m$ at time $t=0$, maximizing
the expected profit $E_{0}[\pi_{F}]$: 
\begin{equation}
E_{0}[\pi_{F}]=\gamma\sqrt{m}-m-c_{F}\label{eq:PiAtteso}
\end{equation}
where $\gamma$ is the expected value of the price and the market
fitness at the time $\tau$ conditional to the available information
at time 0: 
\begin{equation}
\gamma=E_{0}\left[p_{\tau}\theta_{\tau}\right]\label{eq:gamma}
\end{equation}
The effective selling price $p_{\tau}$ is set by market equilibrium,
while the variation of market fitness $\theta_{\tau}$ is determined
by stochastic fitness. The investment level $m$ will be set consistently
with market clearing by maximizing the expected profit given in Eq.~(\ref{eq:PiAtteso})
with respect to $m$:
\begin{equation}
m=\left(\frac{\gamma}{2}\right)^{2}\label{eq:m_best}
\end{equation}
Those quantities will determine the farmer\rq{}s effective profit.

\item \textbf{Consumers}. We model the aggregate demand for food as given
by: 
\begin{equation}
D=\frac{w}{\left(p_{\tau}\right)^{\beta}}\label{eq:Cons}
\end{equation}
where $\beta$ is the demand elasticity. The demand curve indicates
the quantity of the commodity which all consumers taken together would
be willing to buy at each level of price. In the model, the value
of $w$ is a constant parameter.
\item \textbf{Financial Investors} are defined as agents that neither produce
nor consume goods, entering the market at time $\tau-\delta$ very
close to the market clearing ($\delta\ll\tau$). They are informed
of the fitness $\theta_{\tau}$ ($\sim\theta_{\tau-\delta}$) and
are allowed to trade on future contracts written on the produced good
itself and contingent to the realization of $\theta_{\tau}$. While
several investment strategies are in principle possible, we assume
that the investor 
trade contingent goods in a quantity that depends linearly on the
differential of the newly available information $\theta_{0}-\theta_{\tau}$:
\begin{equation}
Q_{S}\left(\theta_{\tau}\right)=\alpha\left(\theta_{0}-\theta_{\tau}\right)\label{eq:Qs}
\end{equation}
where $\alpha$ measures the degree of integration between the financial
and the commodity market. Importantly, we will show that its value
can be determined self consistently if one assumes that the investment
level is chosen in such a way that the average investment return is
maximum.

In case of frictionless infinitely liquid capital markets and synchronous
decisions, $Q_{S}$ can be vanishing either because $\alpha$ tends
to zero (no market integration) or $\tau$ tends to zero (synchronous
agents' decisions, implying no information asymmetry). We will show
that, when information asymmetry and market integration are present,
the net demand $Q_{S}$ of financial investors is zero on average.
Consistently with this condition, we consider the expected capital
gain of the financial investor as: 
\begin{equation}
E_{\tau-\delta}[\pi_{S}]=E_{\tau-\delta}[Q_{S}\ p_{\tau}]-c_{S}\simeq\alpha\left(\theta_{0}-\theta_{\tau}\right)p_{\tau}-c_{S}\label{eq:profit_spec}
\end{equation}
\\
where $c_{S}$ is the average financial transaction cost, $p_{\tau}$
is set by the market equilibrium and $\pi_{S}=\alpha\left(\theta_{0}-\theta_{\tau}\right)p_{\tau}-c_{S}$.
Consistently with the case of the farmer, in Eq.~(\ref{eq:profit_spec})
we do not consider the cost of capital. However, since $\delta\ll\tau$
capital investments are more expensive for farmers than for financial
investors. For this reason, considering the cost of capital in agents' profits will
eventually increase the profit of financial investors, with respect to that of producers. 

\end{itemize}

\subsubsection{The probabilistic description of fitness uncertainty}

\label{2.2} 

Even if all the farmers enter the market with the same fitness $\theta_{0}$,
their fitness will be different at time $\tau$ due to unpredictable
changes in the global production condition of the commodity, affecting
independently each individual farmer. Denoting by $\theta_{\tau}^{i}$
the fitness at time $\tau$ of farmer $i$, we write it as the sum
of normal distributed aggregate and idiosyncratic innovations: 
\begin{equation}
\label{evoltheta}
\theta_{\tau}^{i}=\theta_{0}+\bar{\sigma}W_{\tau}+\sigma W_{\tau}^{i}
\end{equation}
where $W_{\tau}$ and $W_{\tau}^{i}$, $i=1,\ldots,N$ are $N+1$
statistically independent Wiener processes, with $N$ denoting the
number of farmers. $\bar{\sigma}$ and $\sigma$ are positive parameters
that determine the typical size of the variation in the average fitness
and the difference between this average and the fitness of a single
farmer. 

Hereafter, we will assume that the number of farmers is very large,
and denote by $E^{I}[\phi(\theta_{\tau}^{i})]$ the average value
between all the farmers of a function $\phi$ of $\theta_{\tau}^{i}$:
\begin{equation}
E^{I}\left[\phi(\theta_{\tau}^{i})\right]=\lim_{N\rightarrow\infty}\frac{1}{N}\sum_{i=1}^{N}\phi(\theta_{\tau}^{i})=\int_{-\infty}^{\infty}dx\phi(x)N_{x}(\theta_{\tau},\sigma^{2}\tau)\label{eq:idios_ave}
\end{equation}
where $\theta_{\tau}=\theta_{0}+\bar{\sigma}W_{\tau}$ and $N_{x}(a,b)$
is a normal distribution of average $a$ and variance $b.$ 

We will also denote by $E^{\mathbb{A}}[\xi(\theta_{\tau})]$ the average
value of a function $\xi$ of $\theta_{\tau}$ over the realizations
of $W_{\tau}$:

\begin{equation}
E^{\mathbb{A}}[\xi(\theta_{\tau})]=\int_{-\infty}^{\infty}dx\xi(x)N_{x}(\theta_{0},\bar{\sigma}^{2}\tau)\label{eq:avecesar}
\end{equation}

\subsubsection{The time line}
\begin{itemize}
\item At time $0$, the farmers decide the amount of capital to invest in
the production. In order to take this decision, they can access all
the information on the past time course of $\theta$ and of the price.
Based on this information, they know exactly the probability distribution
of the fitness at time $\tau$. Moreover, they know the equations
regulating the commodity market they participate in and anticipate
their investment decision conditional on the realization of the fitness.
\item At time $\tau-\delta$, $\delta\ll\tau$, the approximate value $\theta_{\tau}$
of the aggregate fitness is observed by financial investors who set
their strategy according to Eq. (\ref{eq:Qs}).
\item At time $\tau$ the market for the food commodity clears and the equilibrium
price $p_{\tau}$ is determined, production of non defaulted farmers
is allocated to consumers and profits are distributed to farmers.
For $\alpha \neq 0$, due to commodity and financial market integration, financial investors
with long (short) positions maturing at $\tau$ also contribute to
an extra demand (supply). 
\end{itemize}
\textbf{Farmers' production decision.} The quantity $Q_{\tau}$ brought
to the market at time $\tau$ is conditional on the production decision
taken at time $0$ and on the realization of the fitness (aggregate
plus idiosyncratic). From Eq. (\ref{eq:m_best}) and (\ref{eq:prod}),
we have 
\begin{equation}
Q_{\tau}=\sqrt{m}{\sum_{i}}^{\prime}\theta_{\tau}^{i}=\left(\frac{\gamma}{2}{\sum_{i}}^{\prime}\theta_{\tau}^{i}\right)
\end{equation}
where the primed sum runs over the farmers who are able to make a
positive profit (namely, that do not default). Assuming the number
of farmers is very large, we have: 
\begin{equation}
Q_{\tau}(\theta_{\tau})=\frac{\gamma}{2}E^{I}\left[\theta_{\tau}^{i}\chi_{\left(\theta^{\ast},\infty\right)}\left(\theta_{\tau}^{i}\right)\right]\label{eq:Qf}
\end{equation}
where $\chi_{\left(a,b\right)}$ is the characteristic function of
the interval $\left(a,b\right)$ and $\theta^{\ast}$ is defined as
the minimum value of $\theta$ corresponding to a non negative profit
for the farmers, i.e. $\theta^{\ast}=max\left[0,\frac{1}{p_{\tau}}\left(\frac{\gamma}{2}+c_{F}\frac{2}{\gamma}\right)\right]$
. The value of $\gamma$ entering in $\theta^{\ast}$ is estimated
by Eq.~(\ref{eq:gamma}) as: 
\begin{equation}
\gamma=E^{\mathbb{A}}[p_{\tau}(\theta_{\tau})\ \theta_{\tau}]\label{eq:Ep}
\end{equation}
where $p_{\tau}$ is a function of $\theta_{\tau}$ determined by
the marked clearing (see below) and we have assumed, under rational
expectation, that $E_{0}[\theta_{\tau}]=E^{A}[\theta_{\tau}]$. 

The farmers' total return is computed in a static form as a statistical
average over the farmers and over different realizations of the fitness.
The average farmers' total return, also determined by market clearing
prices, is evaluated considering the average profit $\Pi_{F}$ per
unit of invested capital $M_{F}=m+c_{F}$: 
\begin{equation}
\mu_{F}=\frac{\Pi_{F}}{M_{F}}=\frac{\Pi_{F}}{\left(\frac{\gamma}{2}\right)^{2}+c_{F}}.\label{eq:rF}
\end{equation}
where $m$ has been estimated using Eq.~(\ref{eq:m_best}). The expected
farmers' profit $\Pi_{F}$ is estimated as: 
\begin{equation}
\Pi_{F}=E^{A}\left[E^{I}\left[\pi_{F}(\theta_{\tau}^{i})\right]\right]
\end{equation}
where, putting together Eqs.~(\ref{eq:p1}), (\ref{eq:prod}) and
(\ref{eq:m_best}), $\pi_{F}(\theta_{\tau}^{i})=\frac{\gamma}{2}p_{\tau}\theta_{\tau}^{i}-\frac{\gamma^{2}}{4}-c_{F}$
if $\theta_{\tau}^{i}>\theta^{\ast}$ and $\pi_{F}(\theta_{\tau}^{i})=-\frac{\gamma^{2}}{4}-c_{F}$
otherwise. 

Similarly, the expected fraction of defaulting farmers at time $\tau$
is given by: 
\begin{equation}
f=1-E^{A}\left[E^{I}\left[\chi_{\left(\theta^{\ast},\infty\right)}\left(\theta_{\tau}^{i}\right)\right]\right]\label{eq:frac}
\end{equation}

\textbf{Financial investors' allocation decision.} In this model the
quantity of contingent goods, given in Eq.~(\ref{eq:Qs}), is decided
by the financial investor at time $\tau-\delta$, under the condition
of zero average net supply. This condition is automatically verified,
since $E^{A}[Q_{S}]=\alpha E^{\mathbb{A}}[\theta_{\tau}-\theta_{0}]=0$.
In fact on a single trade the investor can reduce or increase the
amount of production available for consumption, but this deviation
has zero expectation under the farmer information set. Following the
same reasoning used to compute the farmer total return, we interpret
this static condition as a weak form of the dynamic constraint that
financial investors are zero net suppliers of the traded commodity.
Notice that by construction the position held by the financial investor
in the long run will play a role similar to virtual storage with zero contents in average, with
the possibility of both negative and positive inventory fluctuations.

The financial investor may produce a capital gain $\pi_{S}(\theta_{\tau})$,
given in the last term of Eq.~(\ref{eq:profit_spec}), by extracting
the informational rent generated by the possibility to select an optimal
strategy based on the observation of the realized level of fitness.
Since the feasible strategies are market neutral, i.e. on average
the capital invested in the long and short position adds up to zero,
the investment return per unit of dollar long is quantified by: 
\begin{equation}
\mu_{S}=\frac{\Pi_{S}}{M_{S}}\label{eq:rS}
\end{equation}
where $\Pi_{S}$ is determined by the average capital gain accumulated
by the investor over different outcomes of the fitness $\theta_{\tau}$:

\begin{equation}
\Pi_{S}=E^{A}\left[\pi_{S}\left(\theta_{\tau}\right)\right]\label{eq:PL}
\end{equation}
and, abstracting from margin requirements to short sell the commodity,
$M_{S}$ is the expected amount of capital required for the investment
in the long positions:\\
 
\begin{equation}
M_{S}=E^{A}\left[\ \left(Q_{S}(\theta_{\tau})p_{\tau}+c_{S}\right)\ \chi_{(-\infty,\theta_{0})}(\theta_{\tau})\right]
\end{equation}
where $Q_{S}(\theta_{\tau})$ is given by Eq.~(\ref{eq:Qs}) and $c_{S}$
is the average trading transaction cost.

\subsubsection{Commodity Market clearing}

The selling price $p_{\tau}$ and the allocation of resources among
agents are determined by the clearing of the market, namely by the
total supply equaling the total demand, conditioned to the realization
of the fitness uncertainty $\theta_{\tau}$. For $\alpha \neq 0$, in case of an adverse
(favorable) realization, $\theta_{\tau}<\theta_{0}$ ($\theta_{\tau}>\theta_{0}$),
the production supply (consumer demand) is augmented by the extra
supply (demand) generated by the financial agents investing on contingent
goods maturing at $\tau$.

\begin{equation}
\frac{w}{\left(p_{\tau}\right)^{\beta}}=Q_{\tau}(\theta_{\tau})+Q_{S}(\theta_{\tau})\label{eq:Clearing}
\end{equation}
where $Q_{\tau}$ is the quantity of goods produced at time $\tau$
by farmers that have not defaulted and is given by Eq.~(\ref{eq:Qf}),
and $Q_{S}$ is the extra financial investors' demand or supply defined
in Eq.~(\ref{eq:Qs}). This equation must be solved iteratively together
with Eq.~(\ref{eq:Ep}), since in this equation the selling price $p_{\tau}$
as a function of $\theta_{\tau}$ enters explicitly. The existence
of a unique solution for the equilibrium price and the procedure for
solving numerically Eq.~(\ref{eq:Clearing}) together with (\ref{eq:Qf})
is discussed in the Appendix.

\section{Calculation}

\label{calc}

The equilibrium of the model is characterized by solving the market
clearing in Eq.~(\ref{eq:Clearing}). In this manner the consumer
prices $p_{\tau}$, the equilibrium production $Q_{\tau}$ (given
in Eq.~(\protect\ref{eq:Qf})) and the returns on investments $\mu_{S}$
and $\mu_{F}$ (defined in Eqs.~(\protect\ref{eq:rS},\protect\ref{eq:rF})),
as well as other derivative quantities, can be determined for different
values of market integration $\alpha$.

The model's solution depends on the determination of the farmers'
expectation $\gamma$ (given in Eq.~(\ref{eq:Ep})), {which is calculated
using the price schedule $p_{\tau}$ as obtained by solving Eq.~(\ref{eq:Clearing}).
The system of Eq.~(\ref{eq:Ep}) and Eq.~(\ref{eq:Clearing}) cannot
be solved analitically due to the presence of the liquidity constraint
present in the lower extreme of the definite integral appearing in
Eq.~(\ref{eq:frac}). The system is then solved through the numerical
self-consistent iterative procedure is described in the Appendix.
Numerical results will of course depend on the choice of the scenario
parameters, in particular $\beta$, fixing the market elasticity,
and $\bar{\sigma}$ determining fitness volatility. Thus, a comparative
statics of an analytic approximation of the model for the case of
a single farmer has been used to facilitate the setting of the parameters'
values used in the complete model simulation. The analytic approximation
of the model is also given in the Appendix.

The equilibrium solution discussed below has been found for the parameter
set listed in Table~\ref{tab}. In the following we also discuss
the robustness of the results for different choices of the scenario
parameters. 

\section{Results and economic implications}

\label{res}

\subsection{Equilibrium prices}

Figure~{\ref{fig:fig1}} shows the price function }$p_{\tau}\left(\theta\right)$
{solving Eq.~(\ref{eq:Clearing}), for the choice of the scenario
parameters given in Table~\ref{tab}, as a function of the difference
$\theta_{\tau}-\theta_{0}$, for $\alpha=0$ (representing the scenario
with no markets integration) and for a finite value of $\alpha$}.
As expected financial trading has the effect of stabilizing the equilibrium
price by smoothing the price dependence on exogenous shocks. Consistently
with traditional perfect market models, the commodity price volatility
is also reduced as an effect of trading (see Figure~\ref{fig:price}).
According to the model an increased trading efficiency will reduce
price volatility in any analyzed scenario parameter.
We argue that this is probably related to our choice of using statistically independent Wiener processes to
characterize the evolution of the farmers' fitness appearing in Eq.~(\ref{evoltheta}).

\subsection{Return on financial and farm investments}

Expected speculator's and farmer's returns $\mu_{F}$ and $\mu_{S}$
are shown in Figure~\ref{fig:1} as a function of the market integration
parameter $\alpha$. For values of market integration $\alpha$ smaller
than a critical value $\alpha_{c}$, the financial investor's expected
return is negative, mainly due to the finite transaction cost $c_{S}$
considered in the model. We can interpret this result by observing
that for $\alpha<\alpha_{c}$ market integration does not lead to
an advantage for financial investors: i.e. for $\alpha<\alpha_{c}$
the financial transaction costs $c_{S}$ does not allow financial
investors to extract their information rent. However, for larger values
of financial integration the financial investors can exploit their
superior information (see Figure~\ref{fig:1}, top panel). We also
observe that the variance of the financial investor return $\sigma_{S}$
is almost constant in $\alpha$, consistently with the neutral hypothesis
for financial investors' strategies. Notice that in the model it exists
an\ $\alpha^{\ast}$ which maximizes the expected returns for the
financial investors $\mu_{S}$. The financial investor's return depends
on the choice of scenario parameters, whose explicit dependence has
been found in the simplified analytic model of a single farmer (described
in the Appendix). Within the simplified model its dependence on the
volatility of fitness $\bar{\sigma}$ and the market elasticity $\beta$
is shown in Figure~\ref{Fig:rS1} (top). It can be observed that
financial investors' expected return $\mu_{S}$ is higher for highly
risky (high value of $\bar{\sigma}$) and inelastic markets (low value
of $\beta$). The numerical solution for $\mu_{S}$ is also shown
in Figure~\ref{Fig:rS1} (bottom). We observe that numerical and
analytic solutions are qualitatively in agreement in reconstructing
this observable.

Farmer's investment return $\mu_{F}$ is fairly constant and positive
showing a slight maximum in correspondence to $\alpha^{\ast}$ (see
bottom panel in Figure~\ref{fig:1}). In the long run farmer investment
is productive, thus in the absence of liquidity constraints production
is a profitable investment and, using land as a collateral would guarantee
credit availability.

The variance of the farmers' return $\sigma_{F}$ is low in the region
where $\alpha$ is not far from $\alpha_{c}$, while for larger values
of $\alpha$ it grows sharply. The variance $\sigma_{F}$ is a good
indicator of the level of risk faced by a farmer in producing the
real commodity, and the model is able to reproduce the counterintuitive
evidence that the financialization amplifies the variability of returns
on commodity production.

Figure~\ref{fig:returns} quantifies the impact of the farmer's rational
production planning on investment returns $\mu_{S}$ and $\mu_{F}$
by comparing the outcome of a rational farmer which sets the level
of production taking into account the effect of market integration
level $\alpha$, with the outcome of a naive farmer\rq{}s decision
setting production level while ignoring the effect of market integration
($\alpha = 0$) on the expectation value of $\gamma$ (see Eq.~(\ref{eq:Ep})).
The financial investor's return does not change significantly, while
the farmer's return $\mu_{F}$, not surprisingly, shows a sharp reduction
of return as an effect of a suboptimal production planning: naive
farmers pay the maximum rent to financial investors which are better
informed.

Within our model financialization of trade is driven by the asymmetric
impact of market integration. On the one hand the financial investor
increases the investment opportunity and raises the expected return
on investment, while on the other the liquidity constraint which limits
the aggregate investment capacity of the farmer implies that their
return volatility grows sharply with $\alpha$. In other terms the
improved efficiency of production risk sharing exacerbates the effect
of the liquidity constraints which worsen the financing conditions
of real investment and raise the default rates. This is best understood
by analyzing equilibrium production.

\subsection{Equilibrium production risk and the effect of market integration
on production planning}

As the degree of market integration $\alpha$ increases, the production
quantity (shown in Figure~\ref{fig:quantity}) remains fairly constant,
while its volatility increases with $\alpha$. The default rate in
this market is approximately $30\%$ and almost constant with market
integration, while the default rate variance increases with $\alpha$
as a combined effect of the increase in the return volatility and
the fitness fluctuation, $\bar{\sigma}$. Note that the longer the
time to produce the larger the fluctuations in returns from real investment,
thus an efficient subsidization approach should consider a redistribution
of wealth which accounts for the differential levels of liquidity
stress suffered by producers, intermediaries and last minute-traders
with differential \textquotedblleft time to produce\textquotedblright.

To better identify the relation between farmers default and production
volatility $\sigma_{Q}$, in Figure~\ref{fig:varQ} we compare two
different market conditions corresponding to different levels of exogenous
parameters (details reported in the figure caption): one characterized
by a large default rate $f$, the other characterized by a smaller
$f$, in which subsidization effects are simulated by artificially
raising the level of consumers\rq{} demand (which is equivalent
to a softening of the liquidity constraint). As can be seen the financialization
effect, i.e. the relative increase in production volatility with increasing
market integration $\alpha$ is much larger in the market with higher
default rates.

Notice that within our model the main channel of propagation of instability
induced by financialization is the lack of an efficient hedging channel
able to transfer the liquidity risk. Thus, the most dangerous fluctuations
are not price fluctuations but rather those induced by fluctuations
in default rates.

A thorough analysis of the policy
implications under similar assumptions is discussed in \cite{Leathers01111986}, where it is suggested that subsidies
are determined by the necessity to reduce default rates to improve
welfare. Within our equilibrium solution all those conclusions continue
to hold. It is however important to point out an important distinction:
in our model defaults are driven by liquidity rather than credit risk.
This is consistent with 
the well known observation that in most of case subsides are
given to land owners which are well collateralized and have a reduced
exposure to credit risk.

\subsection{Can farmers' hedging be effective? Some basic risk management policy
considerations }

In our model farmers are not allowed to invest in the financial market
and financial investors do not invest in real production. It is important
to notice, however, that this restriction does not impose any limit
in the diversification opportunities of liquidity risk affecting production.
The farmer can reduce the default probability only by reducing the
amount of capital invested in the production of the real commodity.
However, this farmer decision will not have any consequence on aggregate
sector quantities.

Assume that a representative investor is allowed to invest both in
real and financial activities by maximizing the risk return trade-off
on those activities. The asymmetric impact of market integration on
the returns of real production and financial trading would imply that
the optimal allocation in financial investment would increase with
$\alpha$ and the investment in production would be depleted. Changing
the identity of the investors would not modify the clearing equation
(\ref{eq:Clearing}). For this reason the global relations of quantity,
price and farmers' survival fraction (and their volatility) would
remain unaffected by any farmer hedging policy. The key condition
is that the liquidity constraint applies only to real investment and
generates capital market segmentation which is not mitigated by financial
hedging.

Despite our model does not address the welfare implications of the
market imperfections, it suggests that an improvement in the liquidity
risk management should be one of the main goals of optimal market
regulations, taxation schemes and public subsidization programs and
correspondingly that such policies should enhance liquidity risk sharing
between producers, intermediaries and traders investing in commodity
markets. Stated that the longer is the investment horizon, in presence
of large fitness uncertainty, the larger is the liquidity risk, optimal
intervention schemes could be introduced to classify different agents
according to their time to produce.

\section{Conclusions}

\label{conc}

In this paper we have shown that, at least in principle, in commodity
markets dominated by uncertainty and characterized by a \lq\lq{}time
to produce\rq\rq{} $\tau$, at a certain critical level of market
integration a financial trader operating with a deterministic strategy
can take advantage of his \textquotedblleft on time\textquotedblright \ action to
realize a profit. Upon the introduction of a diffusive market stochastic
fitness, we demonstrate that this is the consequence of a competitive
equilibrium generated by the asynchronous decisions of the two agents
participating in market clearing: producers are forced by their production
cycle to anticipate market decisions when programming their production
strategy, while financial investors can instantaneously react to current
prices. In fact, the long production cycle $\tau$ operates as a credit
and liquidity risk multiplier on farmers, which, along with liquidity
constraints on production, makes the long-term (production) investments
much riskier than short-term (financial) investments: due to the presence
of liquidity constraints on long-term producers, the integrated spot commodity
and financial market is inefficient in sharing the risk between market
participants.

The model shows that more \emph{elastic} markets tolerate \emph{larger}
values of market integration. On the contrary negative effects of
market financialization are more pronounced on those markets characterized
by long production cycles and$/$or large market fitness volatility.
Among negative effects we point out that the combined effect of liquidity
constraints and market financialization amplifies producers' risk
and production volatility. This fact may result in markets which are
unprotected by price turbulence if investors move suddenly elsewhere
their financial activity towards different investment targets. The
model presented here can be developed and improved along many directions:
first the above model implications can be tested on real data and
an empirical analysis would certainly improve our understanding of
market imperfections, second our analysis does not take into account
the relevant sustainability issues which are generated by the complex
interaction between commodity production, food quality health programs
and standards of living of consumers. As for the first issue we note
that if spot commodity and financial market integration is proxied by the relative
size of short and long-term allocations, our model predicts that for
a specific commodity this quantity should positively correlate with
production volatility of that commodity.

\newpage{}

\appendix

\section{Supplementary information }

\subsection{The producers' expectation $\gamma$}

The model solution discussed in Section~\ref{mat} depends
on the determination of the farmers' expectation $\gamma=E_{0}[p_{\tau}\theta_{\tau}]$,
which is calculated as given in Eq.~(\ref{eq:Ep}) and using the
price schedule $p_{\tau}$ obtained by solving Eq.~(\ref{eq:Clearing}).
The Eq.~(\ref{eq:Ep}) is a non-linear self consistent equation of
the form $\gamma=F(\gamma)$, where the expectation value $\gamma$
is also present in the second member through the price schedule $p_{\tau}$.
The Eq.~(\ref{eq:Ep}) is solved with respect to $\gamma$ using
a numerical iterative algorithm based on the secant method. The algorithm
starts from a guess $\gamma^{0}$ and find the value that satisfy
the convergence criteria %
\mbox{%
$|\gamma^{N}-F(\gamma^{N})|<10^{-7}$%
}, after $N$ iterations.

At each iteration $n$ the second member $F(\gamma^{n})$ is calculated
by computing the integral numerically using a grid of $133$ points
over the range of possible $\theta_{\tau}$. Even though the distribution
of $\theta_{\tau}$ has an unbounded support, in practice we truncate
the distribution at four standard deviation from the mean $\theta_{0}$.
The number of grid points has been chosen in such a way that the integral
$F(\gamma^{N})$ does not change more than $1\%$ between two consecutive
grid points. The integrand $p_{\tau}^{n}(\theta_{\tau})\ \theta_{\tau}$
is evaluated at each value of $\theta_{\tau}$ on the grid and a constant
interpolation between the grid points is used for values of $\theta_{\tau}$
not on the grid. The integrand evaluation is obtained by solving numerically
the clearing equation Eq.~(\ref{eq:Clearing}) in the form: 
\begin{equation}
p_{\tau}^{n}=\left(\frac{w}{Q_{\tau}^{n}(\theta_{\tau})+Q_{S}(\theta_{\tau})}\right)^{\frac{1}{\beta}}
\end{equation}
where $Q_{\tau}^{n}(\theta_{\tau})$ is given in Eq.~(\ref{eq:Qf})
and is a function $Q(p_{\tau}^{n},\theta_{\tau},\gamma^{n})$, while
$Q_{S}$ is given in Eq.~(\ref{eq:Qs}) and is a function of $\theta_{\tau}$
only. The numerical iterative algorithm starts from a guess value
$p_{\tau}^{n,0}$ and finds the equilibrium price when the convergence
criteria $|p_{\tau}^{n,M}-p_{\tau}^{n,M-1}|<10^{-7}$ over $M$ iterations
is satisfied. In practice convergence is achieved after $M\times N$
iterations, where $M$ is typically between $20$ and $30$, and $N$
between $10$ and $10^{3}$ depending on the initial guess values
for $\gamma^{0}$.

The clearing equation Eq.~(\ref{eq:Clearing}) has a single solution
as demonstrated in Section~A.2 . The existence of a single solution
for the expectation $\gamma$ can be analytically demonstrated in
the case of a single farmer (see Section~A.3). In the more general
case, we checked that the algorithm described above converges always
to the same value by starting the solution search from different initial
guess values for $\gamma^{0}$, distributed at random between $0$
and $10000$.

\subsection{Existence condition for the equilibrium price}

For a given realization of the fitness uncertainty $\theta_{\tau}$,
the demand curve in Eq.~(\ref{eq:Clearing}) is a monotonic decreasing
function of $p_{\tau}$, while, due to the price dependence of $Q_{\tau}$
through $\theta^{\ast}$ (see Eq.~(\ref{eq:Qf})), the supply curve
is monotonic increasing in $p_{\tau}$. For this reason the price
forms, i.e. it exists the equilibrium solution of Eq.~(\ref{eq:Clearing}),
if $Q_{\tau}+Q_{S}>0$. Since $Q_{\tau}$ is limited from above the
existence condition becomes: 
\begin{equation}
\frac{\gamma}{2}\int_{0}^{\infty}dxN_{x}\left(\theta_{\tau},\sigma^{2}\tau\right)x\geq\alpha\left(\theta_{\tau}-\theta_{0}\right)\label{eq:existCond}
\end{equation}
that also depends on the financial investor's strategy. For $\alpha=0$
the solution always exists. When this condition is satisfied the clearing
equation always has a single positive solution $p_{\tau}$.

\subsection{Simplified equilibrium solution for a single farmer}

In this section we provide the analytic solution of a simplified formulation
of the commodity market model described in this paper. We consider
a single farmer case. Thus, Eq.~(\ref{eq:Clearing}) becomes: 
\begin{equation}
\frac{w}{\left(p_{\tau}\right)^{\beta}}=\frac{\gamma}{2}\theta_{\tau}-\alpha\left(\theta_{\tau}-\theta_{0}\right)\label{eq:Clearing_simp}
\end{equation}

Under the hypothesis that, in order to make his$/$her production
decision, the producer takes the price schedule $p_{\tau}(\theta)$
solving the clearing equation (\ref{eq:Clearing_simp}) with $\alpha=0$:
\begin{equation}
p_{\tau}(\theta)=\left(\frac{2w}{\gamma\theta}\right)^{\frac{1}{\beta}}
\end{equation}
the expected value $\gamma=E_{0}[p_{\tau}\theta_{\tau}]$ solving
Eq.~(\ref{eq:Ep}) reduces to: 
\begin{equation}
\gamma=\int d\theta\ N_{\theta}(\theta_{0},\bar{\sigma}^{2}\tau)\left(\frac{2w}{\gamma}\right)^{\frac{1}{\beta}}\theta^{1-\frac{1}{\beta}}\label{eq:Gamma}
\end{equation}
Developing the probability distribution of $\theta_{\tau}$ in cumulants,
Eq.~(\ref{eq:Gamma}) can be solved with respect to $\gamma$ and
we obtain the single solution: 
\begin{equation}
\gamma=(2w)^{\frac{1}{1+\beta}}\theta_{0}^{\frac{\beta-1}{\beta+1}}\left[1-\frac{1}{2\beta}(1-\frac{1}{\beta})x^{2}\right]^{\frac{\beta}{\beta+1}}\label{eq:gamma1}
\end{equation}
where $x=\frac{\sigma}{\theta_{0}}$.

Expanding $\gamma$ into a power series with center $x=0$, the constant
($\gamma_{0}$), 1st-order ($\gamma_{1}$) and 2nd-order ($\gamma_{2}$)
approximation are respectively: 
\[
\gamma_{0}=\left(2w\right)^{\frac{1}{1+\beta}}\theta_{0}^{\frac{\beta-1}{\beta+1}}\gamma_{1}=0\gamma_{2}=-\gamma_{0}\frac{\beta-1}{\beta(\beta+1)}x^{2}
\]

The approximate analytic form of the financial investor's return (Eq.~(\ref{eq:rS}))
is then: 
\begin{equation}
\mu_{S}=\frac{Ax^{2}-c_{S}}{Bx-Ax^{2}+c_{S}}\label{eq:approx}
\end{equation}
where: 
\begin{equation}
A=\frac{\alpha\gamma_{0}}{\beta}-\frac{2\alpha^{2}}{\beta}B=\frac{\alpha\gamma_{0}}{\sqrt{2\pi}}
\end{equation}
and $c_{S}$ is the financial transaction cost.

\newpage{}

\newpage

\renewcommand\thetable{\arabic{table}}
\setcounter{table}{0}
\begin{table}[!h]
\begin{minipage}[c]{0.9\textwidth}%
\centering
\begin{tabular}{|c|c|c|c|c|c|c|}
\hline 
$\beta$  & $w$  & $\bar{\sigma}$  & $\sigma$  & $\theta_{0}$  & $c_{F}$  & $c_{S}$ \\
\hline 
0.6  & 0.02  & 0.1  & 0.2  & 0.5  & 0.6  & 0.0002 \\
\hline 
\end{tabular}
\vspace{0.3cm}
\caption{The chosen set of external parameters used in the model. They have
been set in order to satisfy the following requirements: the equilibrium
existence condition (Eq.~(\ref{eq:existCond})), the existence of
a range of values for the market integration $\alpha$ for which the
financial investor's expected return is positive, and the requirement
that the effect on the system induced by the presence of financial
investors is not negligible. There exist different sets of parameters
satisfying those requirements, however, as discussed in Section~3
the results are robust for different choices of the external parameters.
The value of $c_{F}$ is in units of the optimal investment
level $m$ set in Eq.~(\ref{eq:m_best}).}
\label{tab} 
\end{minipage}
\end{table}

\newpage{}

\renewcommand\thefigure{\arabic{figure}}
\setcounter{figure}{0}
\begin{figure}[!h]
 \centering \includegraphics[width=0.8\linewidth,natwidth=789,natheight=593]{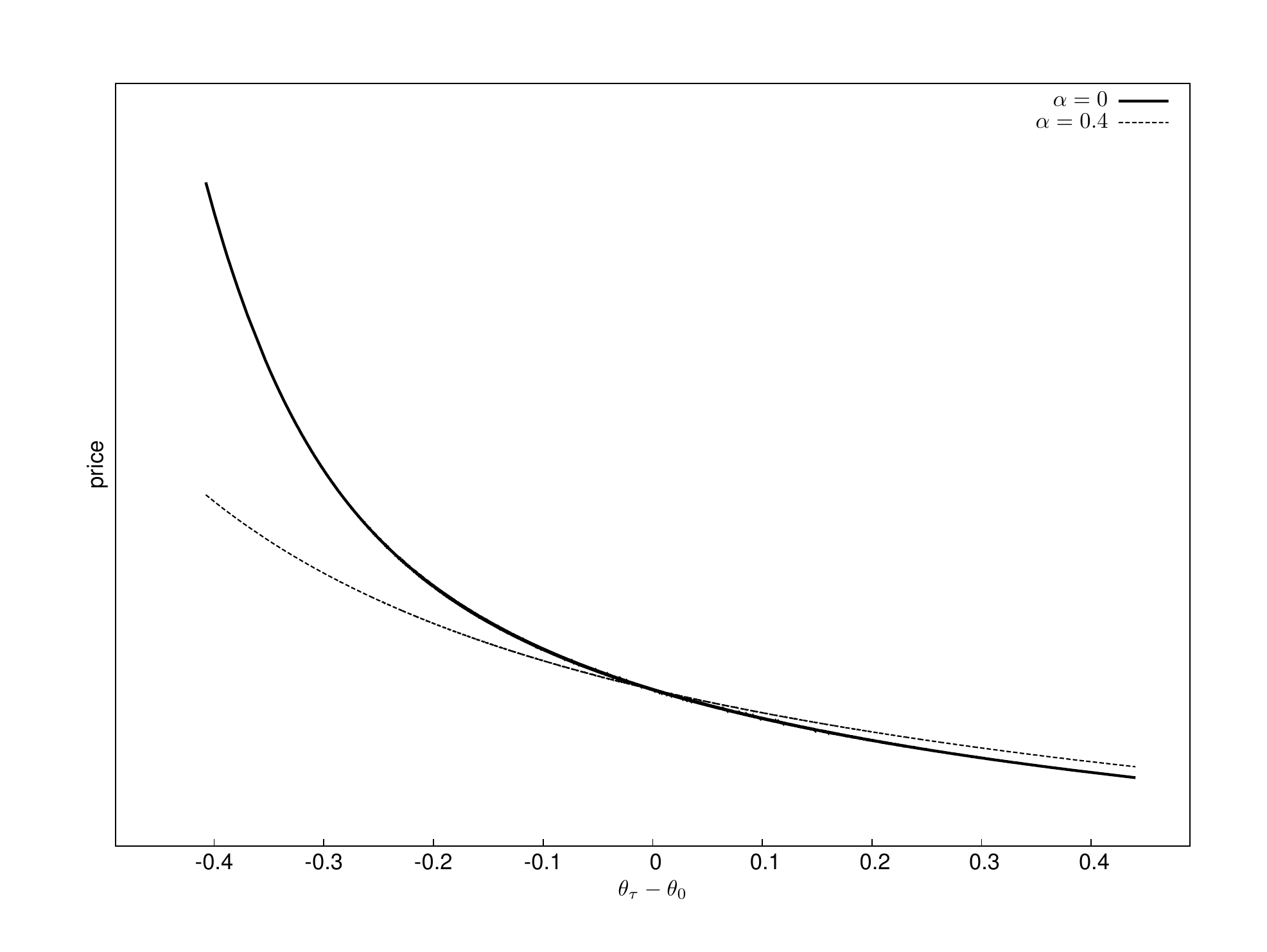}
\protect\caption{The equilibrium price, derived from the solution of the equation (\protect\ref{eq:Clearing}),
as a function of the difference $\theta_{\tau}-\theta_{0}$ for $\alpha=0$
(no financial and commodity markets integration) and for a finite
value of $\alpha$ ($\alpha=0.4$). The set of parameters that have
been used is provided in Table~\ref{tab}.}
\label{fig:fig1} %
\end{figure}

\newpage{}

\begin{figure}[p]
 \centering 
 \includegraphics[width=0.8\linewidth,natwidth=789,natheight=593]{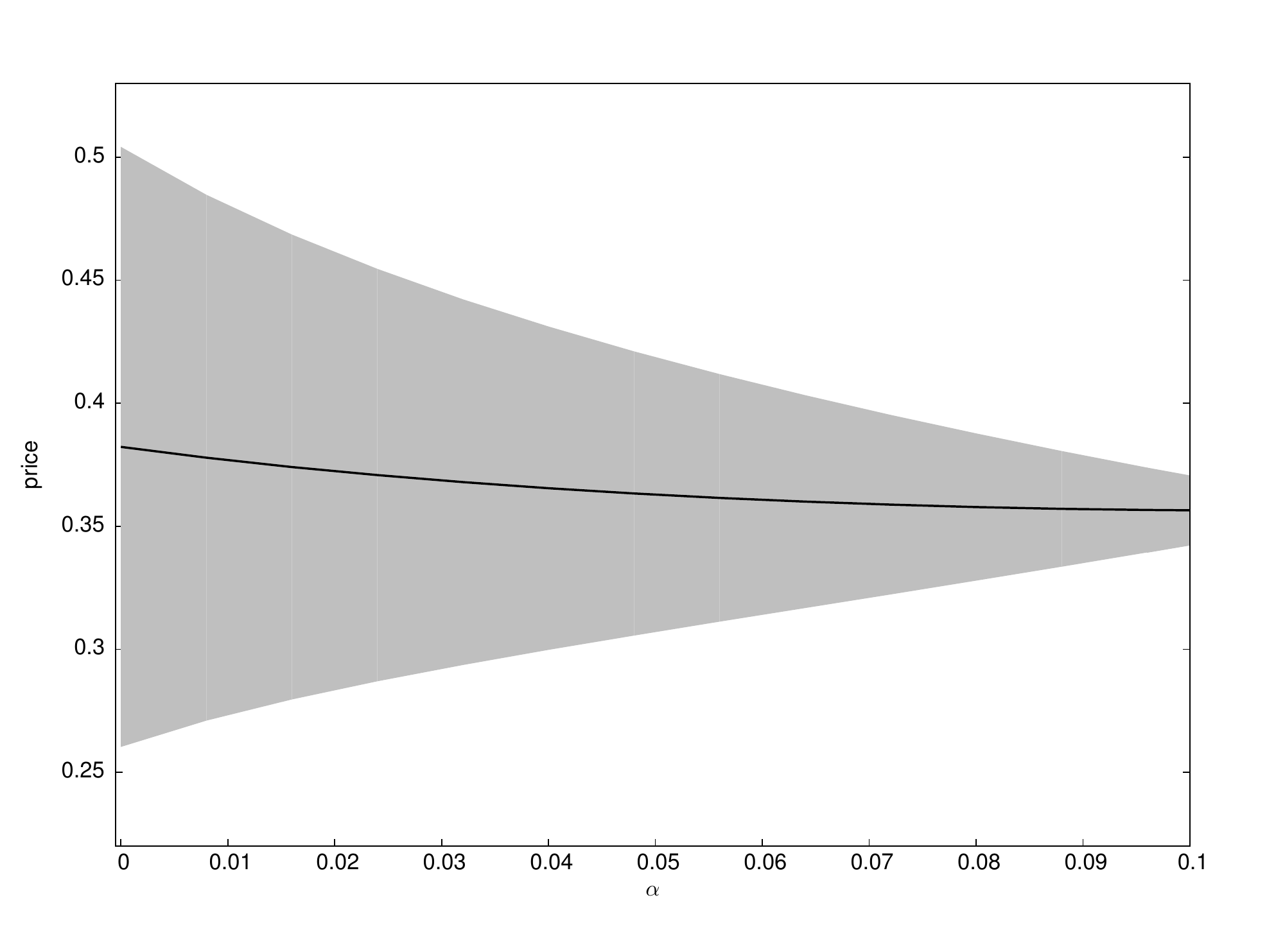}
\protect
\caption{The average equilibrium price $p_{\tau}$ as obtained by solving Eq.~(\protect\ref{eq:Clearing})
is shown as a function of $\alpha$. Prices are averaged over different
realizations of $\theta_{\tau}$ with distribution function $N_{\theta_{\tau}}(\theta_{0},\bar{\sigma}^{2}\tau)$
(see Section \ref{2.2}). The supplied quantity $Q_{\tau}(\theta_{\tau})$
appearing in Eq.~(\protect\ref{eq:Clearing}) is averaged over not
defaulted farmers (see Eq.~(\ref{eq:Qf})). The error bars correspond
to the $1\sigma$ variation.}
\label{fig:price} %
\end{figure}

\newpage{}

\begin{figure}[!h]
 \centering \includegraphics[width=0.8\linewidth,natwidth=789,natheight=789]{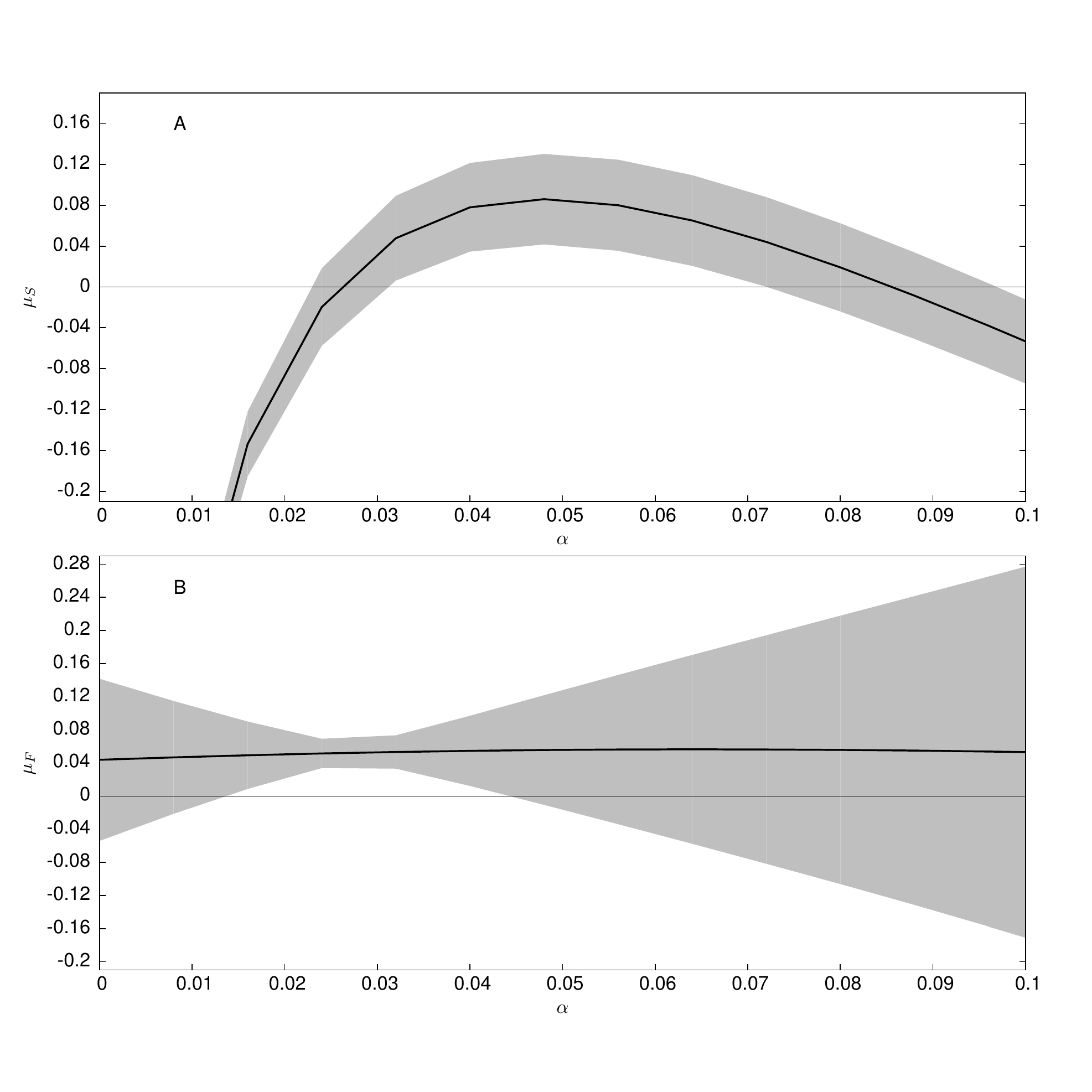}
\protect
\caption{The average returns $\mu_{S}$ (Eq.~(\protect\ref{eq:rS})) of the
speculator and of the farmers $\mu_{F}$ (Eq.~(\protect\ref{eq:rF}))
are shown as a function of $\alpha$. $\mu_{S}$ is averaged over
different realizations of $\theta_{\tau}$ with distribution function
$N_{\theta_{\tau}}(\theta_{0},\bar{\sigma}^{2}\tau)$ (see Section~\ref{2.2}),
while $\mu_{F}$ is also averaged over the farmers\rq{} population
$N$. The error bars correspond to the $1\sigma$ variation.}
\label{fig:1} %
\end{figure}

\newpage{}

\begin{figure}[!h]
 \centering \includegraphics[width=0.8\linewidth,natwidth=789,natheight=593]{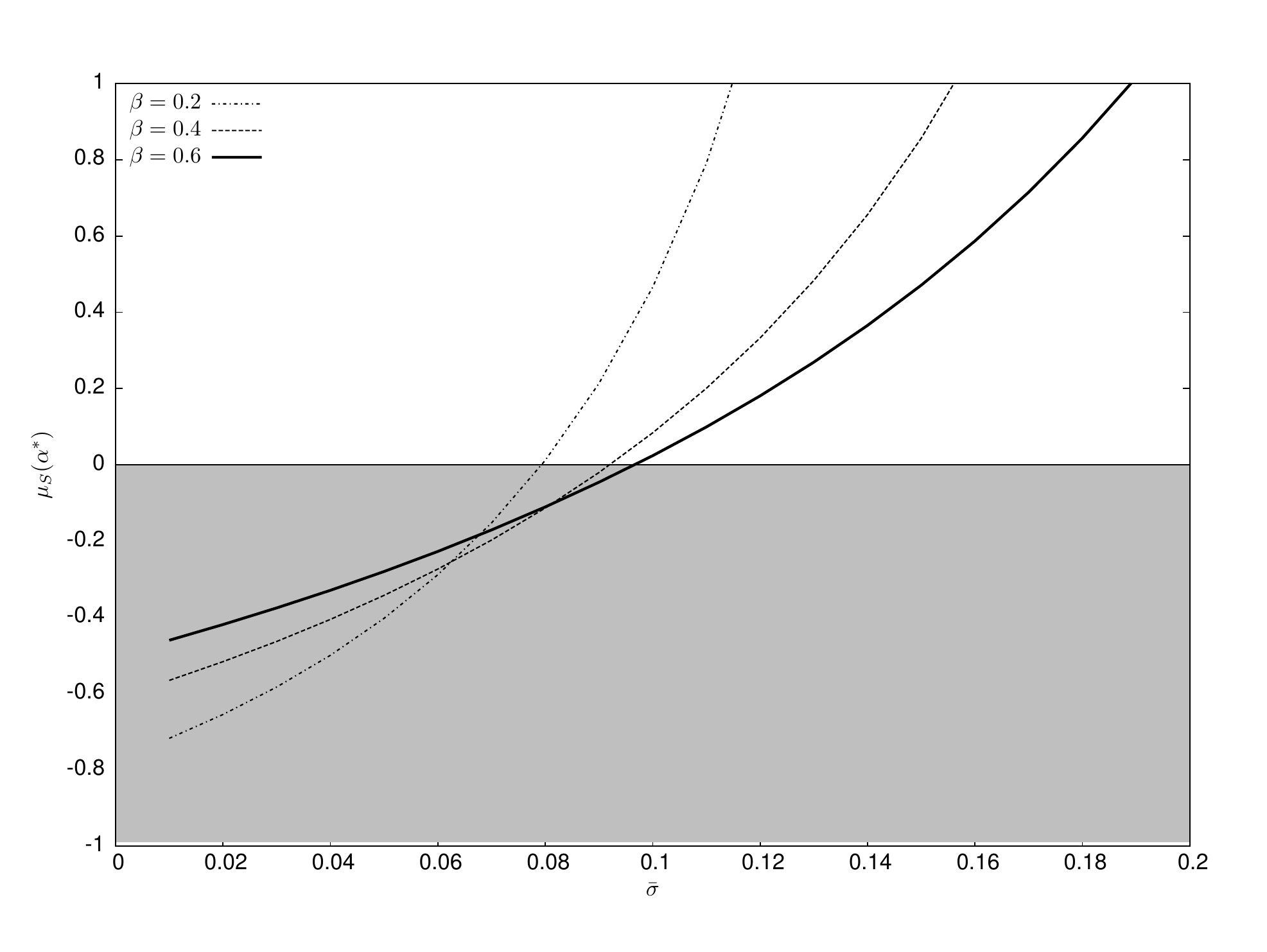}\\
 \includegraphics[width=0.8\linewidth,natwidth=789,natheight=593]{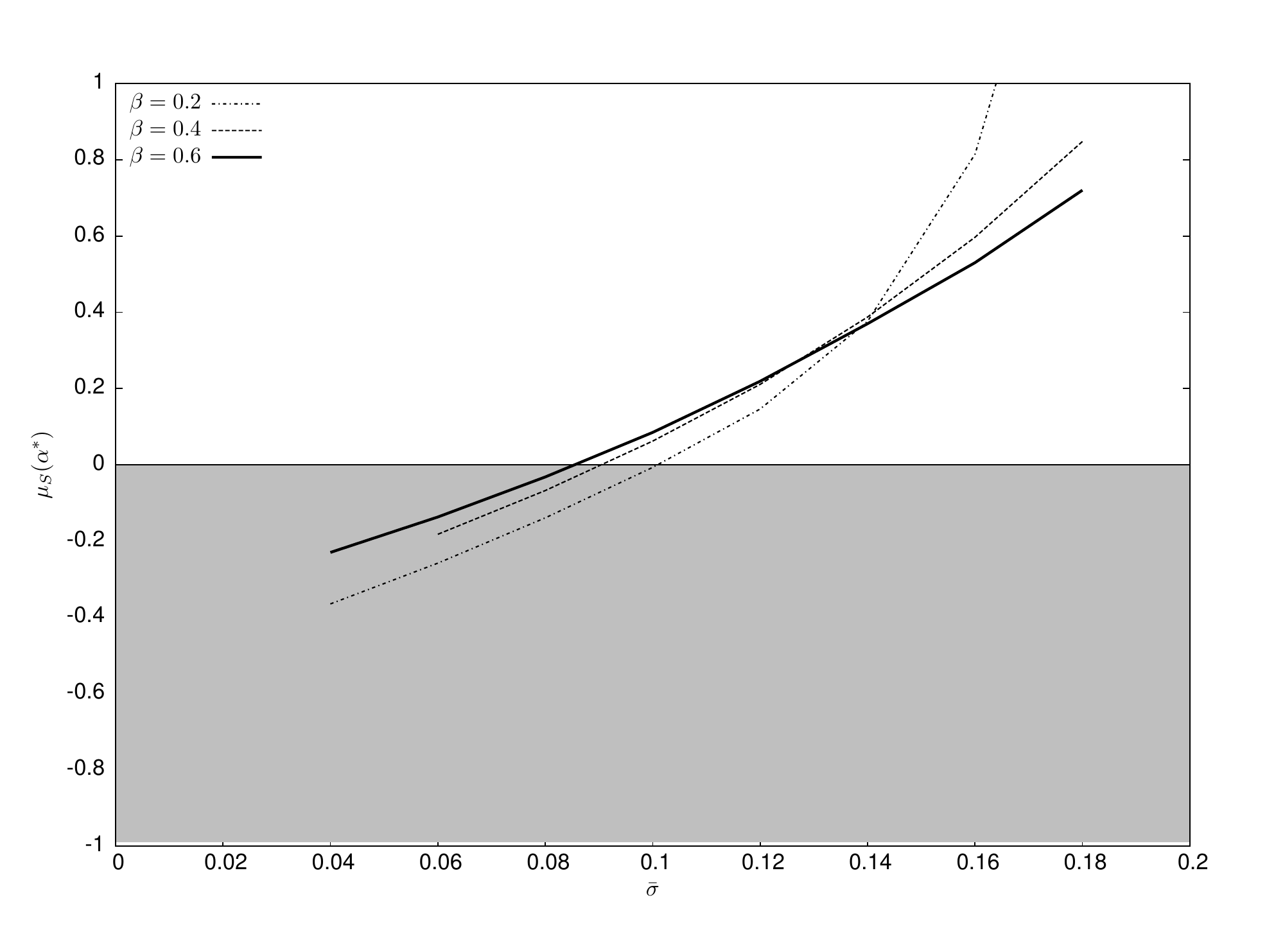}
\protect\caption{The financial investor's return $\mu_{S}$ evaluated at the ``optimum''
$\alpha^{\ast}$ is shown as a function of the externalities' dispersion
$\bar{\sigma}$, and three different values of the market elasticity
$\beta$: $\beta=0.2$ (dot dashed line), $\beta=0.4$ (dashed line) and $\beta=0.6$
(solid line). On the top: results from the analytic calculation shown in
Appendix A. At the bottom: average results from the numerical calculation.
Averages of $\mu_{S}$ are calculated as specified in the caption
of Figure~\ref{fig:1}. }
\label{Fig:rS1} %
\end{figure}

\newpage{}

\begin{figure}[!h]
 \centering \includegraphics[width=0.8\linewidth,natwidth=789,natheight=789]{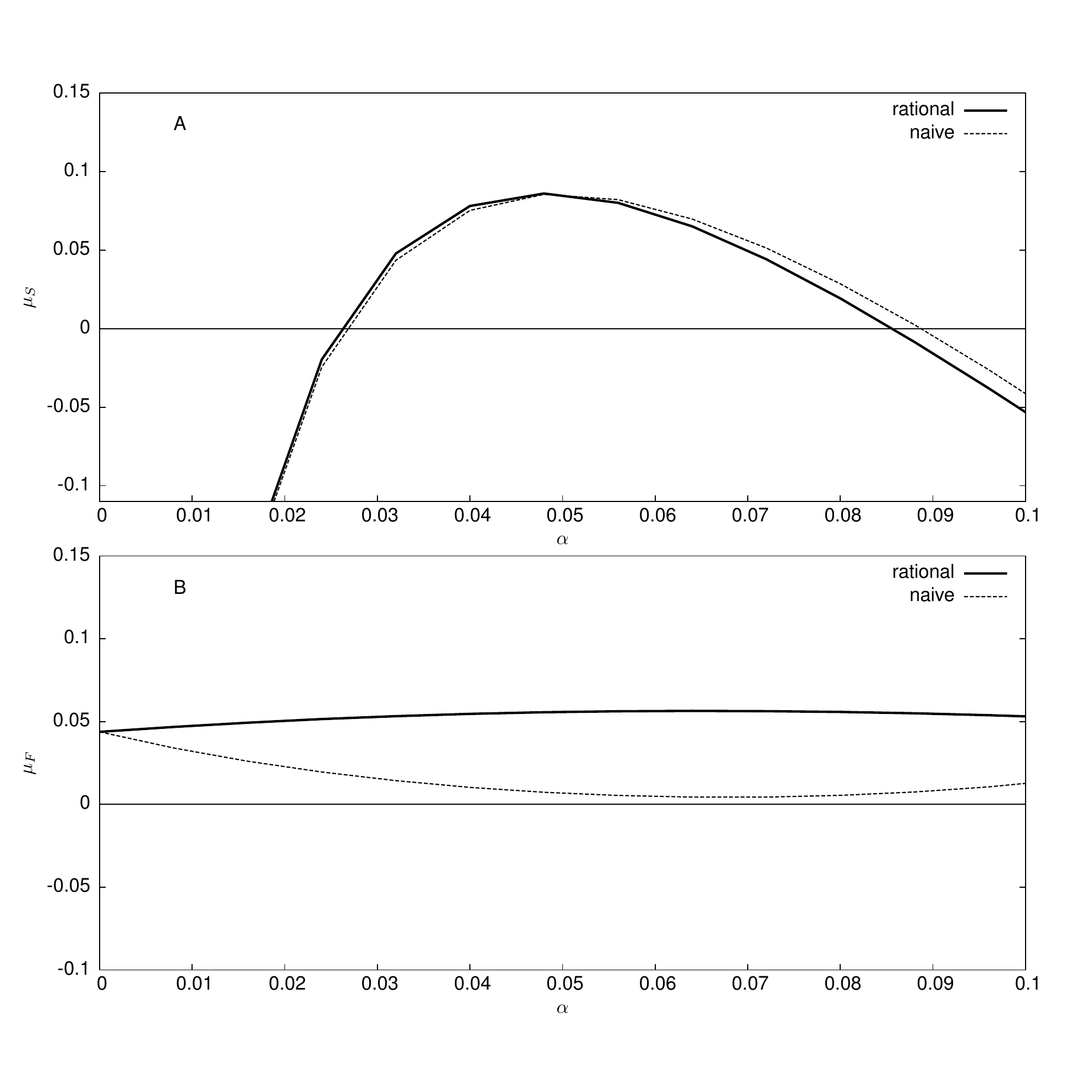}
\protect
\caption{The average returns $\mu_{S}$ (Eq.~(\protect\ref{eq:rS})) of the
speculator and of the farmers $\mu_{F}$ (Eq.~(\protect\ref{eq:rF}))
are shown as a function of $\alpha$. Results for the rational and
naive hypothesis for the farmers are compared. Averages are calculated
as specified in the caption of Figure~\ref{fig:1}. }
\label{fig:returns} %
\end{figure}

\newpage{}

\begin{figure}[!h]
\centering %
 \includegraphics[width=0.8\linewidth,natwidth=801,natheight=789]{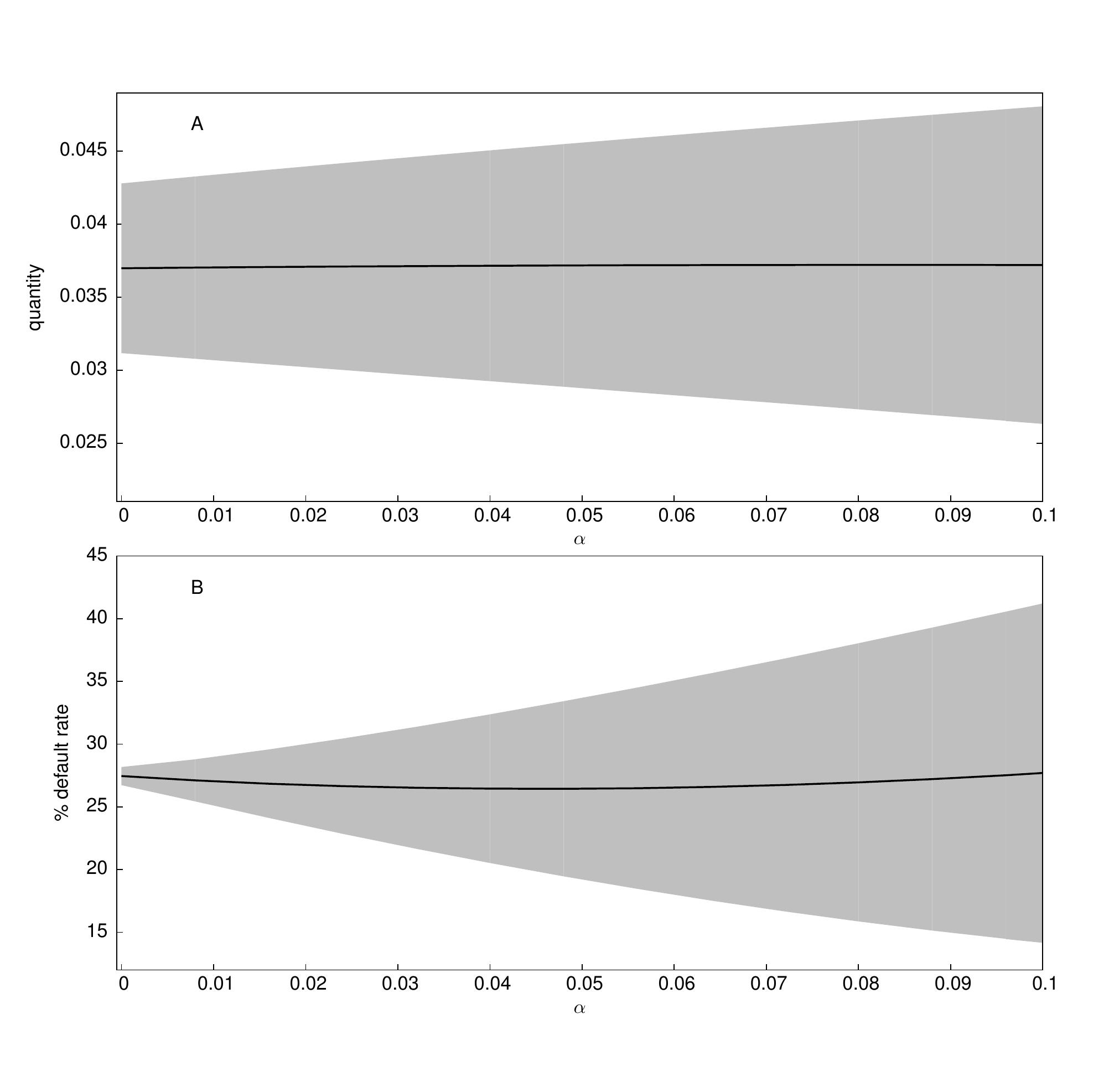}
\protect\caption{ The aggregate quantity produced $Q_{\tau}$ given in Eq.~(\protect\ref{eq:Qf})
and the fraction $f$ of farmers that default given in Eq.~(\protect\ref{eq:frac})
are shown as a function of $\alpha$. The fraction $f$ and the supplied
quantity $Q_{\tau}$ are averaged over different realizations of $\theta_{\tau}$
with distribution function $N_{\theta_{\tau}}(\theta_{0},\bar{\sigma}^{2}\tau)$
(see Section \ref{2.2}), and over not defaulted farmers (see Eq.~(\ref{eq:Qf})).
The error bars correspond to the $1\sigma$ variation.}
\label{fig:quantity} %
\end{figure}

\newpage{}

\begin{figure}[!h]
 \centering \includegraphics[width=0.8\linewidth,natwidth=821,natheight=593]{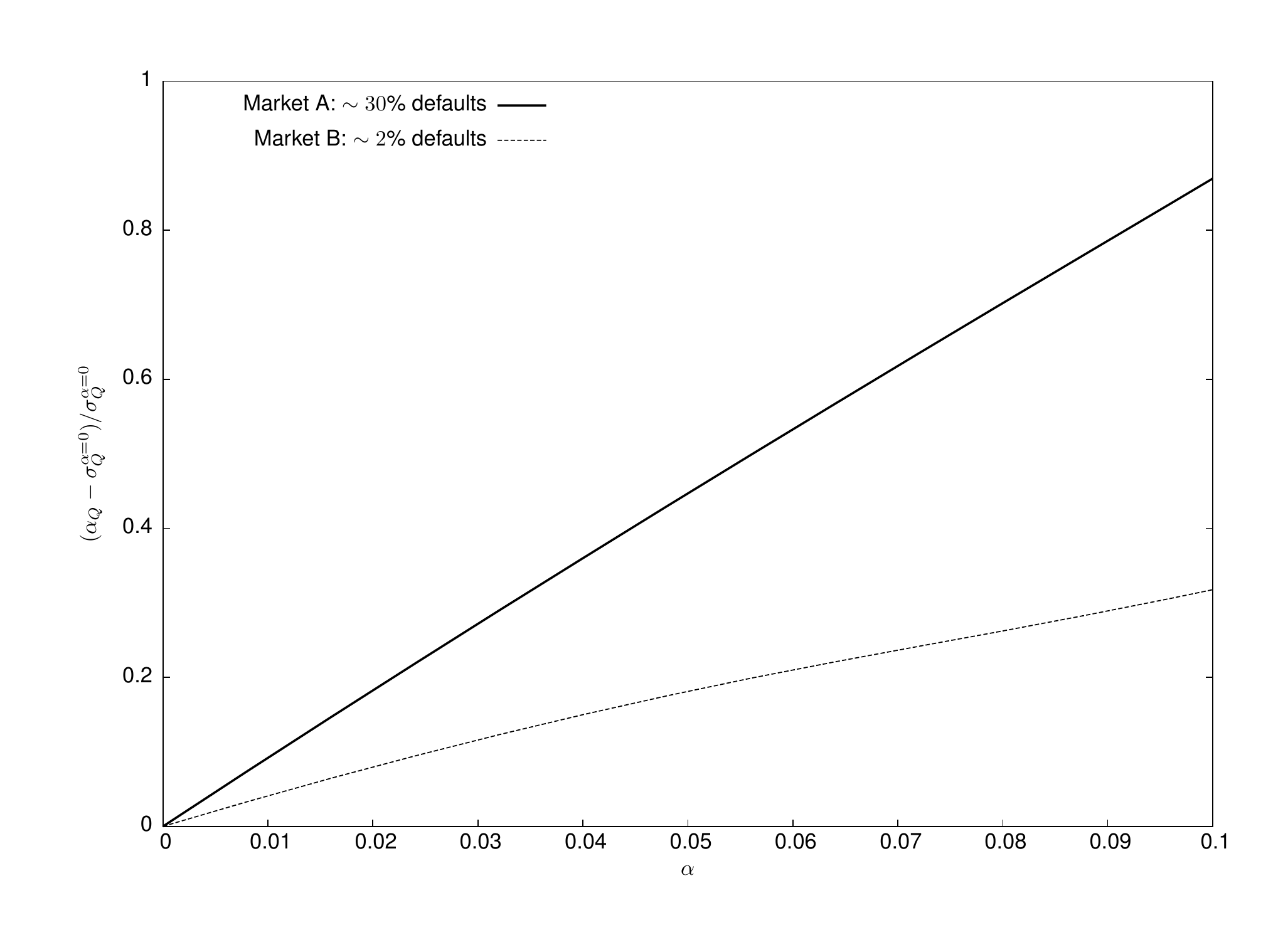}
\caption{The variation of production volatility as a function of the market
integration. Solid line: Market~A with high fraction of farmer defaults.
Dashed line: Market~B with low fraction of farmer defaults. Market~A
is characterized by the set of parameters reported in Table~1. The
set of parameters characterizing Market~B differs for a larger fitness
uncertainty $\bar{\sigma}$ and a higher consumers level of demand
(increased $w$). This market results in a significantly smaller farmer
defaults fraction ($\sim$$2\%$, to be compared to $\sim$$30\%$
of defaults in Market~A, see Section 3.3).}
\label{fig:varQ} 
\end{figure}


\begin{thebibliography}{10}
\expandafter\ifx\csname url\endcsname\relax
  \def\url#1{\texttt{#1}}\fi
\expandafter\ifx\csname urlprefix\endcsname\relax\def\urlprefix{URL }\fi
\expandafter\ifx\csname href\endcsname\relax
  \def\href#1#2{#2} \def\path#1{#1}\fi

\bibitem{Orhangazi2008}
O.~Orhangazi, {Financialisation and capital accumulation in the non-financial
  corporate sector: A theoretical and empirical investigation on the US
  economy: $1973-2003$}, Cambridge Journal of Economics 32~(6) (2008) 863--886.

\bibitem{Soros2008}
G.~Soros, Testimony at the Senate Hearing On Energy Market Manipulation and
  Federal Enforcement Regimes. URL:
  http://www.gpo.gov/fdsys/pkg/CHRG-110shrg80428/html/CHRG-110shrg80428.htm
  (2008).

\bibitem{Tadesse2013}
G.~Tadesse, B.~Algieri, M.~Kalkuhl, J.~von Braun,
  \href{http://dx.doi.org/10.1016/j.foodpol.2013.08.014}{{Drivers and triggers
  of international food price spikes and volatility }}, Food Policy 47 (2014)
  117--128.
\newline\urlprefix\url{http://dx.doi.org/10.1016/j.foodpol.2013.08.014}

\bibitem{Mello2013}
A.~S. Mello, J.~E. Parsons, {Margins, Liquidity, and the Cost of Hedging},
  Journal of Applied Corporate Finance 25~(1) (2013) 34--43.

\bibitem{Parsons2011}
J.~E. Parsons, A.~S. Mello,
{{Rising food prices: What hedging can and cannot do}},
  Blog: Betting the Business Financial risk management for non-financial
  corporations. URL: http://bettingthebusiness.com/
  2011/06/27/rising-food-prices-what-hedging-can-and-cannot-do/ (2011) [cited
  2013].

\bibitem{Cooper1999}
I.~A. Cooper, A.~S. Mello, {Corporate Hedging: the Relevance of Contract
  Specifications and Banking Relationships}, European Finance Review 2 (1999)
  195--223.

\bibitem{Allayannis1999}
G.~Allayannis, {Comment on "Corporate Hedging: The Relevance of Contracts
  Specifications and Banking Relationships"}, European Finance Review 2 (1999)
  225--228.

\bibitem{Mello2000}
A.~S. Mello, J.~E. Parsons, {Hedging and Liquidity}, Review of Financial
  Studies 13~(1) (2000) 127--153.

\bibitem{Leathers01111986}
H.~D. Leathers, J.-P. Chavas,
  \href{http://ajae.oxfordjournals.org/content/68/4/828.abstract}{Farm debt,
  default, and foreclosure: An economic rationale for policy action}, American
  Journal of Agricultural Economics 68~(4) (1986) 828--837.
\newblock \href
  {http://arxiv.org/abs/http://ajae.oxfordjournals.org/content/68/4/828.full.p%
df+html}
  {\path{arXiv:http://ajae.oxfordjournals.org/content/68/4/828.full.pdf+html}},
  \href {http://dx.doi.org/10.2307/1242129} {\path{doi:10.2307/1242129}}.

\bibitem{Brennan1958}
M.~J. Brennan, The supply of storage, American Economic Review 48 (1958)
  50--72.

\bibitem{Williams1991}
J.~C. Williams, B.~D. Wright, {Storage and Commodity Markets}, Cambridge:
  Cambridge University Press, 1991.

\bibitem{Deaton-Laroque}
A.~Deaton, G.~Laroque, {On the Behavior of Commodity Prices}, Review of
  Economic Studies 59 (1992) 1--23.

\bibitem{Deaton1996}
A.~Deaton, G.~Laroque, Competitive storage and commodity price dynamics,
  Journal of Political Economy 104~(5) (1996) 896--923.

\bibitem{Routledge2000}
B.~R. Routledge, D.~J. Seppi, C.~S. Spatt, Equilibrium forward curves for
  commodities, Journal of Finance 60 (2000) 1297--1338.

\bibitem{Keynes1930}
J.~M. Keynes, {A Treatise on Money: The Applied Theory of Money}, Vol.~2,
  London: Macmillan, 1930.

\bibitem{Hicks1939}
C.~Hicks, {Value and Capital}, New York: Oxford University Press, 1939.

\bibitem{Hirshleifer1988}
D.~Hirshleifer, {Residual Risk, Trading Costs, and Commodity Futures Risk
  Premia}, Review of Financial Studies 1~(2) (1988) 173--193.

\bibitem{Samuelson1971}
P.~A. Samuelson, Stochastic speculative price, Proceedings of the National
  Academy of Science 68 (1971) 335--337.

\bibitem{Griggeman-Towe-Morehart}
B.~C. Briggeman, C.~A. Towe, M.~J. Morehart,
  \href{http://ideas.repec.org/a/oup/ajagec/v91y2009i1p275-289.html}{{Credit
  Constraints: Their Existence, Determinants, and Implications for U.S. Farm
  and Nonfarm Sole Proprietorships}}, American Journal of Agricultural
  Economics 91~(1) (2009) 275--289.
\newline\urlprefix\url{http://ideas.repec.org/a/oup/ajagec/v91y2009i1p275-289.%
html}

\bibitem{Chavas2000}
J.-P. Chavas, {On information and market dynamics: The case of the U.S. beef
  market}, Journal of Economic Dynamics \& Control 24 (2000) 833--853.

\bibitem{Vercammen}
J.~Vercammen,
  \href{http://EconPapers.repec.org/RePEc:oup:erevae:v:34:y:2007:i:4:p:479-500%
}{Farm bankruptcy risk as a link between direct payments and agricultural
  investment}, European Review of Agricultural Economics 34~(4) (2007)
  479--500.

\bibitem{Zhou1998}
Z.~Zhou, {An Equilibrium Analysis of Hedging with Liquidity Constraints,
  Speculation, and Government Price Subsidy in a Commodity Market}, The Journal
  of Finance 53~(5) (1998) 1705--1736.

\bibitem{Acharya2009}
V.~Acharya, L.~Lochstoer, T.~Ramadorai, {Limits to Arbitrage and Hedging:
  Evidence from Commodity Markets}, Working paper, New York University.

\end{thebibliography}
\end{document}